\documentclass[letter]{aa}

\usepackage[utf8]{inputenc}
\usepackage{graphicx}
\usepackage[export]{adjustbox}
\usepackage{stackengine}
\usepackage{epstopdf}
\usepackage{url}
\usepackage{natbib}
\usepackage{xcolor}
\usepackage{multirow}
\usepackage{gensymb}
\usepackage{subcaption}
\usepackage{float}
\usepackage[percent]{overpic}
\usepackage{array}
\usepackage{booktabs}
\usepackage[switch]{lineno}

\newcommand\CHECK[1]{\textcolor{red}}
\newcommand{\IGR}{IGR\,J11014--6103}
\newcommand{\igr}{IGR\,J11014--6103~}
\newcommand{\lneb}{Lighthouse~Nebula~}
\newcommand{\LNEB}{Lighthouse~Nebula}

\begin{document}

    \title{Signatures of extended radio emission from escaping electrons in the Lighthouse Nebula}
    \titlerunning {Radio emission from the Lighthouse Nebula}
    \author{ P. Bordas\inst{1}, X. Zhang \inst{1}, V. Bosch-Ramon\inst{1} \and J.M. Paredes\inst{1} }
    \institute{Departament de Física Quàntica i Astrofísica, Institut de Ciències del Cosmos, Universitat de Barcelona, IEEC-UB, Martí i Franquès 1, 08028, Barcelona, Spain }
    \date{Received --- ; accepted ---}

%%%%%%%%%%%%%%%%%%%%%%%%%%%%%%%%%%%%%%%%%%%%%%%%%%%%%%%%%%%%%%%%%%%%%%%%%%%%%%%%%%%%%%%%%%%%%%%%%%%

\abstract{Several supersonic runaway pulsar wind nebulae (sPWNe) with jet-like extended structures have been recently discovered in X-rays. If these structures are the product of electrons escaping the system and diffusing into the surrounding interstellar medium, they can produce a radio halo extending for several arcmins around the source. We model the expected radio emission in this scenario in the Lighthouse Nebula sPWN. We assume a constant particle injection rate during the source lifetime, and isotropic diffusion into the surrounding medium. Our predictions strongly depend on the low- and high-energy cutoffs given in the particle distribution. Our results indicate that extended radio emission can be detected from the \lneb without the need to invoke extreme values for the model parameters. We provide synthetic synchrotron maps that can be used to constrain these results with observations by current highly sensitive radio instruments.}

\maketitle
%\linenumbers

%%%%%%%%%%%%%%%%%%%%%%%%%%%%%%%%%%%%%%%%%%%%%%%%%%%%%%%%%%%%%%%%%%%%%%%%%%%%%%%%%%%%%%%%%%%%%%%%%%%

                                \section{\IGR, the Lighthouse Nebula}
                            \label{section:introduction}
%           ************************************************************

Pulsar wind nebulae (PWNe) interacting with the surrounding interstellar medium (ISM) can give rise to distinct extended morphological features. Torus-like structures and bipolar jets are observed in  pulsars displaying slow proper motions, as well as bow-shaped shocks and extended cometary tails in pulsars moving at high velocities (see e.g. \citealp{Gaensler2006}). In the X-ray band, these structures have been resolved in detail in a number of cases (\citealp{Kargaltsev2008}). A small subset of supersonically moving PWN (sPWN from now on) have, in addition, been observed to display unusually long X-ray outflows. Some of the most prominent examples of such outflows are found in the Guitar Nebula \citep{Hui2012} and in the Lighthouse Nebula \citep{Pavan2014}, but the number of large-scale filaments from sPWNe is gradually increasing, see for example the recent findings in PSR~J1135-6055 \citep{Bordas2020}, J1809--1917 \citep{Klingler2020}, PSR~J2030+4415 \citep{Vries2020}, PSR J2055+2539 \citep{Marelli2019}, or the filaments found in the Galactic Center region such as G0.13-0.11 \citep{Zhang2020}, likely associated to a sPWN (see also previous findings in \citealp{Kargaltsev2017}). The origin of these extended jet-like structures is unknown, but their length, almost rectilinear geometry, and misaligned orientation with respect to the pulsar proper motion are difficult to be accommodated for in the framework of canonical pulsar jet formation theories, where jets are produced by the magnetic collimation of the plasma downstream of the pulsar wind termination shock (see e.g. \citealp{Lyubarsky2002, Komissarov2004, Porth2014} and references therein).
This led to alternative interpretations, including a scenario in which particles accelerated at the pulsar wind/medium shock interface escape the PWN, and diffused into the ambient medium magnetic field, yielding synchrotron emission which is then observed in the X-ray band as quasi-rectilinear jet-like structures (\citealp{Bandiera2008}). Recent numerical simulations seem to support such a scenario, explaining also the asymmetric morphologies observed at opposite sides of the pulsar \citep{Barkov2019a, Bucciantini2020}. 

In this paper we model the radio emission produced by high-energy electrons escaping the sPWN \IGR, the Lighthouse Nebula, into the ISM, where they radiate in the ISM magnetic field. After a brief description of the source, in Section \ref{section:model} we outline the model used to describe the emitting electron population and their diffusion into the ISM. Section \ref{section:results} provides particle and radiation distributions, as well as synthetic maps at radio wavelengths used to predict the expected signal for dedicated observations of the source by highly sensitive facilities. Section \ref{section:discussion} discusses the potential of radio observations to constrain the origin of the jet-like structures observed from sPWNe.

%======================================
% %________________________________________________________________________
% %\begin{comment}
% \begin{figure}[t]
% %
% \centering
% \includegraphics{FIGURES/LN_sketch_v3.pdf}
% caption{
% Chandra/ACIS exposure-corrected 0.5 -- 7.0 keV image of IGR J11014–6103 (the “Lighthouse nebula”).The image has a pixel size of 0.492$\arcsec$ and has been smoothed with a 2D-Gaussian kernel with a radius of 2.2$\arcsec$.
%   }
%  \label{figure:sketch}
% \end{figure}

%_____________________________________________________________________

\begin{figure}[h]
  \centering
    \includegraphics[width=0.5\textwidth]{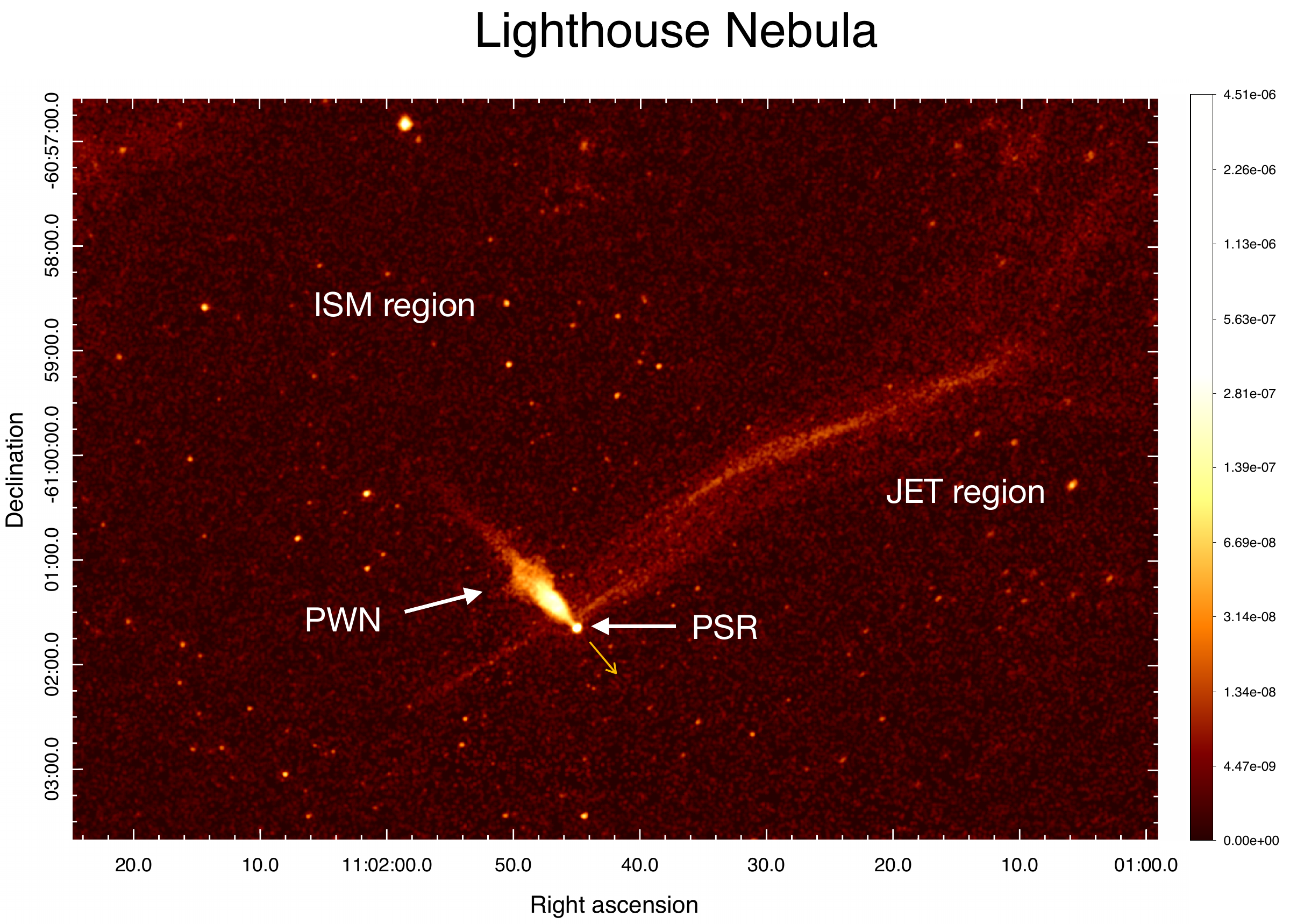}
  \caption{\textit{Chandra}/ACIS exposure-corrected broadband (0.5 -- 7.0 keV) flux image of IGR J11014–6103 (the ``Lighthouse Nebula''). The image has been produced using all \textit{Chandra} archival observations of the \lneb, for a total of $\sim365$~ks, merged using CIAO's {\texttt{merge\_obs}} tool, and smoothed using a 2D Gaussian kernel with radius  on a pixel size of 0.492$\arcsec$ and has been smoothed with a 2D-Gaussian kernel with radius = 1.7$\arcsec$ and $\sigma = 0.7$\arcsec.}
  \label{figure:sketch}
\end{figure}
%======================================

Originally discovered by {\em INTEGRAL} as an unidentified hard X-ray source, \igr (also known as the ``Lighthouse Nebula''; \citealp{Pavan2014}) is located $\sim$11 arcmins southwest of its likely progenitor supernova remnant SNR MSH 11-61A, at a distance of $\sim$7 kpc from the Sun (see \citealp{Garcia2012}). The \lneb is powered by the pulsar PSR~J1101$-$6101, displaying a period of $P = 62.8$~ms and an estimated spin-down energy of $L_{\rm sd} = 1.36 \times 10^{36}~\mathrm{erg\ s^{-1}}$ (\citealp{Halpern2014}). Observations in the X-ray band have revealed a complex system morphology, including a trailing PWN, shaped in a narrow cone elongated toward the parent SNR, and prominent X-ray jet-like structures oriented nearly perpendicular to the PWN axis (see \citealp{Pavan2016}). At radio energies, extended emission has been detected with ATCA at 2~GHz from the \igr tail \citep{Pavan2014}. This emission is not spatially coincident with that observed at X-rays, with the peak of the radio emission being separated further away along the tail by about 22$\arcsec$ with respect to the X-ray maximum.

The jets observed in the \lneb  display a rectilinear structure extending for more than 10 pc into the surrounding medium (see Fig.~\ref{figure:sketch}). Their X-ray luminosity is unusually high, at the level of that retrieved from the pulsar and the tail structures, altogether representing a few percent of the pulsar spin-down power, $L_{\rm sd} \sim 10^{36}$~erg~s$^{-1}$. The jet X-ray spectrum is moderately hard, with photon indices $\Gamma_{\rm X, jets} \approx 1.6$, whereas the tail region displays $\Gamma_{\rm X, tail} \approx 1.9$. Both the jet spectral hardness and its high X-ray efficiency may be expected if these are magnetic jets carrying a hard electron distribution, and there is an amplified magnetic field along the jet structures (see e.g. \citealp{Bykov2017}). However, accounting for the high-speed velocity of the source, $v_{\rm psr} \gtrsim 1\,000~\mathrm{km~s^{-1}}$, the absence of any significant bending along their $\gtrsim 10$~pc is striking prompts for alternative mechanisms for the production of such jet-like structures. Other sPWNe have been interpreted in a diffusion scenario (\citealp{Bandiera2008}), and rectilinear structures or ``filaments'' have also been recently reported from the Galactic Center region (\citealp{MeerKAT2018}), and discussed as ``kinetic jets'' of escaping particles diffusing into the ambient magnetic field (\citealp{Barkov2019b}). It is currently uncertain whether the jet-like structures in the \lneb are ballistic jets carrying an amplified magnetic field or filaments of escaping high-energy electrons from the sPWN into the ISM.

            \section{Modeling the radio emission in a diffusion scenario}
                                \label{section:model}
%           ************************************************************         

The pulsar PSR J1101-6101 in the \lneb displays a spin-down power of $1.36 \times 10^{36}$~erg/s \citep{Halpern2014}. Observations with the \textit{Chandra} satellite report a 2--10~keV X-ray flux from the jets at the level of 6$\times 10^{-13}$~erg~cm$^{-2}$~s$^{-1}$, whereas spectral analysis suggests a synchrotron origin rather than a thermal emission scenario \citep{Pavan2014}. Particles accelerated at the Lighthouse Nebula that are responsible for this emission could also emit in the radio band. 
Under these premises, we consider the presence of a nonthermal population of relativistic electrons injected close to the sPWN bow shock, carrying a total kinetic luminosity $L_{\rm inj} \approx L_{\rm X} \times \, t_{\rm sync}/t_{\rm dyn}$. Here $L_{\rm X}$ is the jet X-ray luminosity derived from \textit{Chandra} observations (we assume a distance to the Lighthouse Nebula of $d_{\rm LN} = 7$~kpc), $t_{\rm sync}$ is the synchrotron cooling time for electrons emitting at the characteristic frequency $h\nu_{\rm c} = 2$~keV in a given magnetic field $B$, $t_{\rm sync} \approx 460\,(B/30 \rm{ \mu G})^{-1/2}$~yr, and $t_{\rm dyn}$ is the dynamical timescale to travel along the jet-like feature, $t_{\rm dyn} \sim l_{\rm jet}/v_{\rm dyn}$; with $l_{\rm jet} \sim 10$~pc and $v_{\rm dyn} \sim c$, $t_{\rm dyn} \sim 32$~yr. A total power $L_{\rm inj} = 3 \times 10^{34}$~erg~s$^{-1}$ is therefore injected, which is distributed in energy following a power-law with index $\Gamma = 2$ and an exponential cutoff at both the low- and the high-energy ends of the particle spectrum at $\gamma_{\rm min}^{\rm cut}$ and $\gamma_{\rm max}^{\rm cut}$, respectively. A single power-law model has been assumed for simplicity in our model, although we note that PWNe could inject a more complex particle distribution displaying a spectral break following the presence of separate electron populations, for example those injected within the pulsar's light cylinder, and electrons accelerated at the shock between the pulsar wind and the external medium (see e.g. \citealp{Atoyan1996, Venter2007, Torres2014}). The low-energy cutoff is an assumption in our model, and it is treated as a free parameter, with two cases explored in this report:$\gamma_{\rm min}^{ \rm cut} = 10^3$ and $10^{5}$. Such a low-energy cutoff may be expected in case an energy selection mechanism is in place in \igr for particles able to escape the PWN. We use $\gamma_{\rm max} = 10^{8}$ for the high-energy particle distribution cutoff, which should be attainable in the bow shock of a runaway pulsar such as \igr (compatible with the jet X-rays; see e.g. \citealp{Bykov2017}). As for the spectral index, we note that the value $\Gamma = 2$ is on the other hand adopted for simplicity, but also because it predicts potentially detectable fluxes. If the electron distribution were much harder, then little emission would be produced in the radio band. 

%======================================

%________________________________________________________________________
\begin{figure}[t]
\centering
\includegraphics[width=0.5\textwidth]{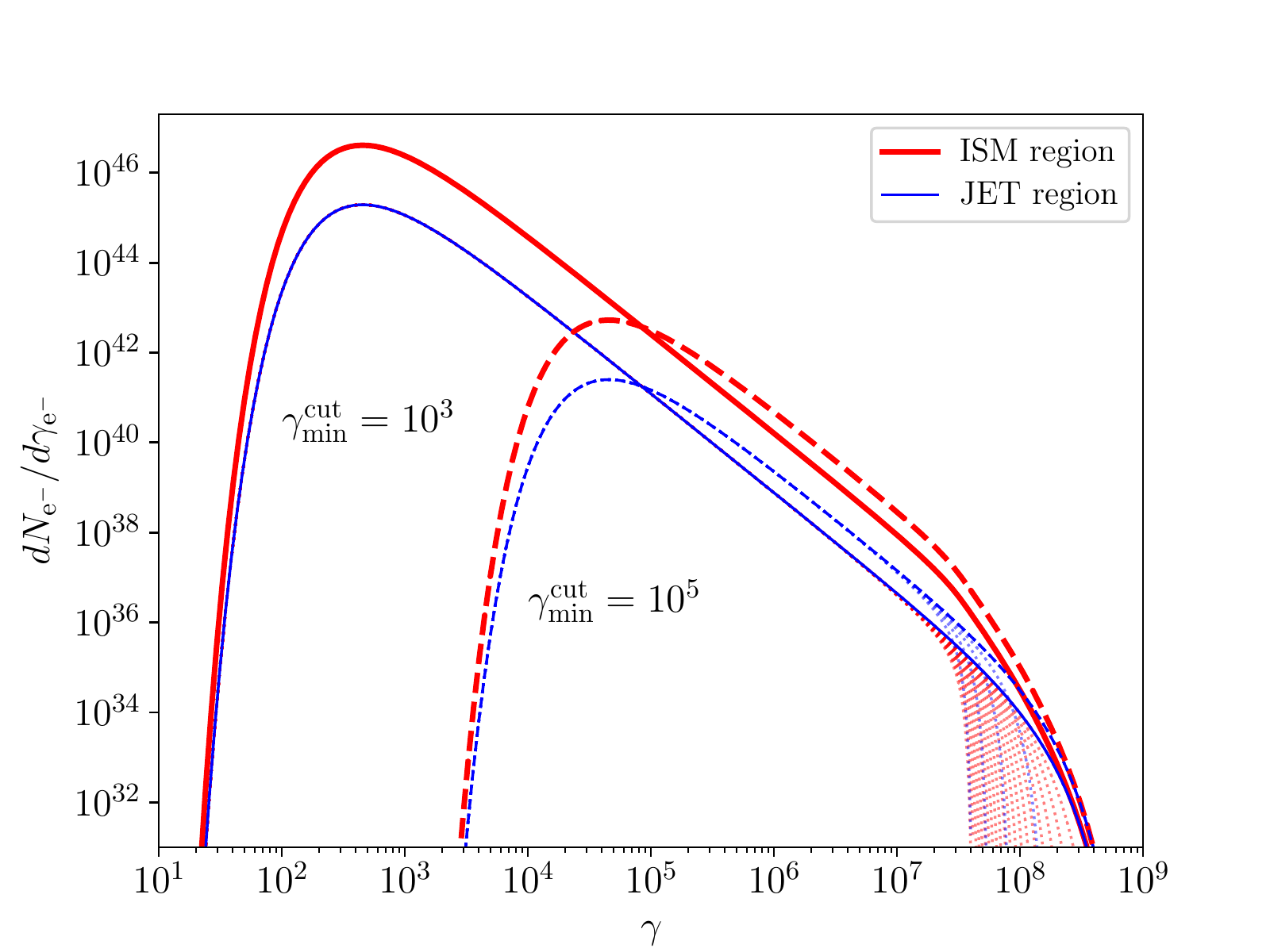}
\caption{Energy distribution of non-thermal particles injected into the JET and ISM regions for different assumed values of the low-energy cutoff $\gamma_{\rm min}^{cut}$. Particle injection is discretized as occurring every $t_{\rm inj} = 1$~kyr (dotted lines). The final cumulative distribution at present time accounts for radiation cooling due to IC on the CMB and synchrotron losses. Distributions are shown for particles injected recently ($t_{\rm inj} \leq 1$~kyr) that diffuse through the JET region (blue lines), and particles injected at $t_{\rm inj} > 1$~kyr diffusing into the ISM region (in red lines). In both cases, a magnetic field of $B_{\rm ISM} = B_{\rm JET} = 5\mu$G is assumed. A low cutoff for the particle distribution at injection is considered, adopting two different values $\gamma_{\rm min}^{cut} = 10^{3}$  and $\gamma_{\rm min}^{cut} = 10^{5}$ (solid and dashed lines, respectively). A high energy cutoff $\gamma_{\rm max}^{cut} = 10^{8}$ is also assumed in all cases. The distribution normalization is computed accounting for the source observed X-ray luminosity (see text for details).}
\label{figure:particle_distribution}
\end{figure}
%________________________________________________________________________
%======================================

%======================================
%________________________________________________________________________
%\begin{comment}
\begin{figure*}[t]
 \begin{subfigure}{.5\textwidth}
   \centering
   \includegraphics[width=0.95\textwidth]{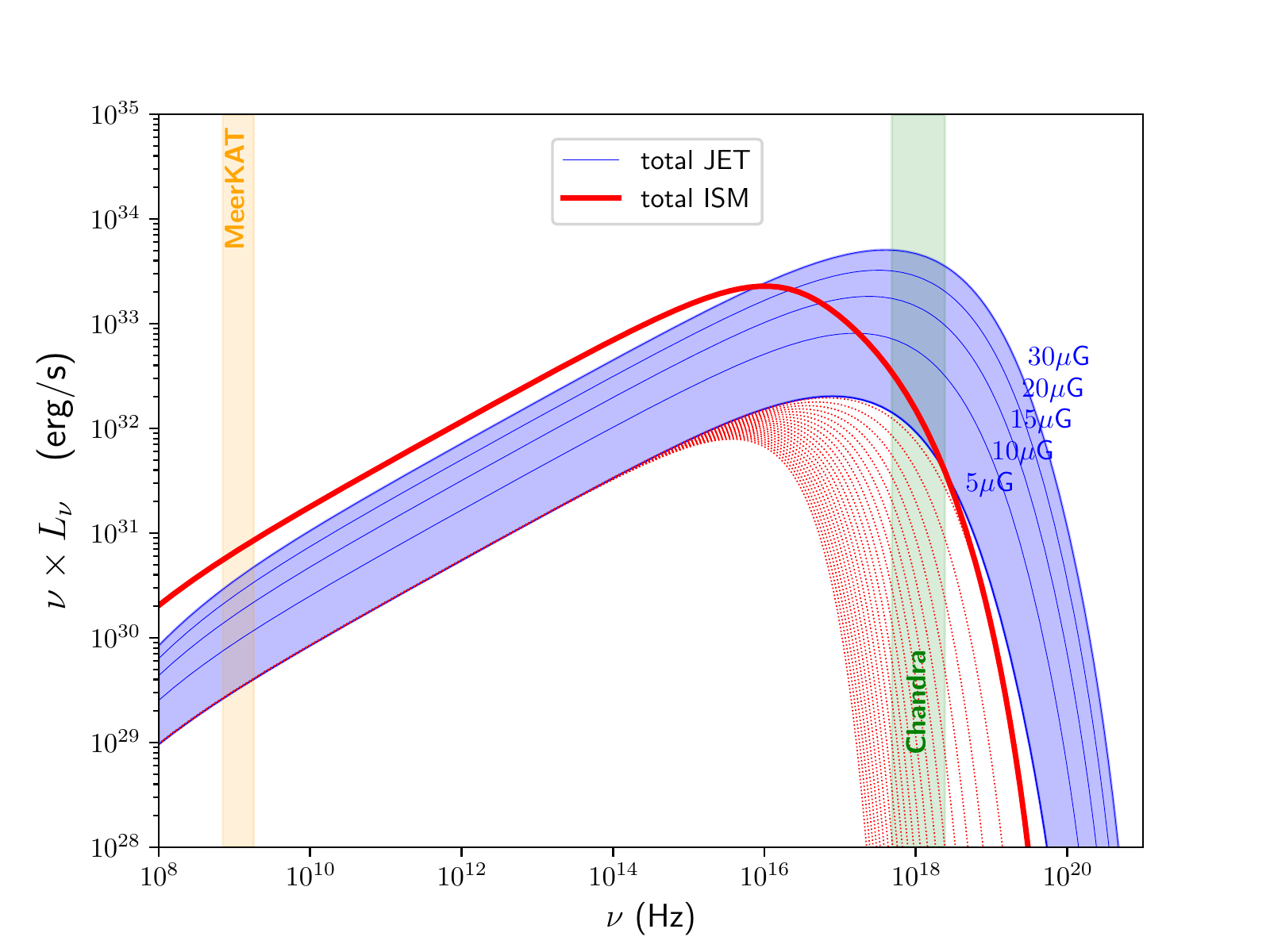}
   \end{subfigure}%
 \begin{subfigure}{.5\textwidth}
   \centering
   \includegraphics[width=0.95\textwidth]{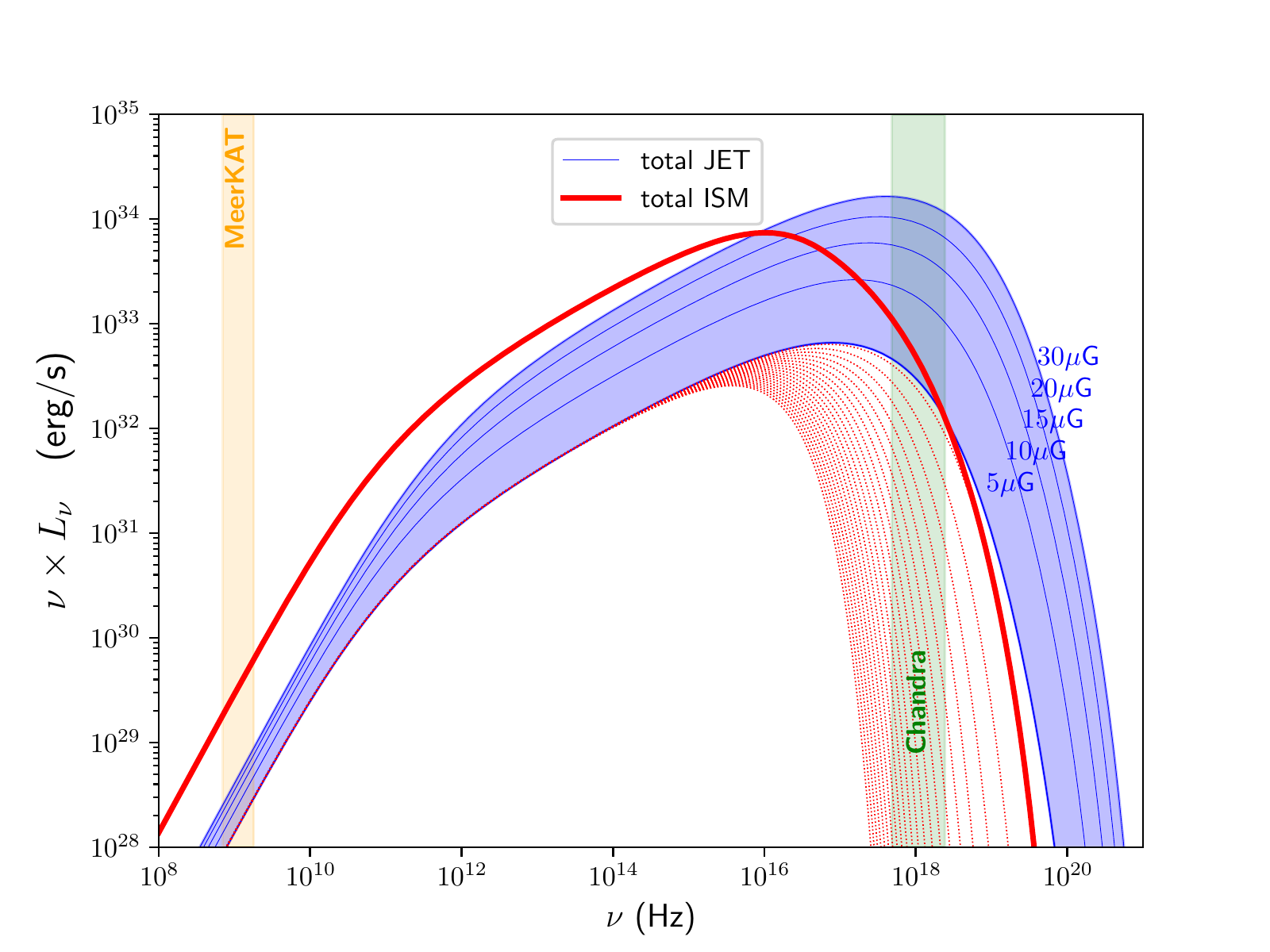}
   \end{subfigure}
  \caption{
  \textbf{Left}: Spectral energy distribution of the emission produced by electrons injected in the ISM (in red) and JET regions (in blue). A magnetic field $B_{\rm ISM} = 5\mu$G is assumed for the ISM, whereas a range in the value of $B_{\rm JET} = 5, 10, 15, 20$ and $30\mu$G is probed for the JET region. Particle injection is considered up to $t_{\rm inj} = 10^3$~yr for the JET region and from $t_{\rm inj} = 10^3$~yr until $t_{\rm inj} = 20$~kyr for the ISM region (red-dotted lines). Particles are assumed to follow a power-law distribution with a low and a high energy cutoff $\gamma_{\rm min}^{cut} = 10^{3}$ and $\gamma_{\rm max}^{cut} = 10^{8}$, respectively. Green and orange vertical bands denote \textit{Chandra} and MeerKAT frequency coverage. \textbf{Right}: Same as in the left panel but for a particle spectrum with a low-energy cutoff $\gamma_{\rm min}^{cut} = 10^{5}$.
  }
 \label{figure:SEDs}
\end{figure*}
%\end{comment}
%_____________________________________________________________________
%======================================

Particles injected close to the bow shock are assumed to escape the immediate sPWN region and diffuse into the surrounding medium, where they evolve with time due to radiation losses, both synchrotron and inverse Compton (IC) emission, the latter accounting for the surrounding cosmic microwave background (CMB) and infrared radiation fields ($U_{\rm CMB} \approx 0.25$~eV~cm$^{-3}$, $U_{\rm IR} \approx 0.2$~eV~cm$^{-3}$, \citealp{Popescu2017}). We evaluate the resulting cooled particle distribution at the present time, $t_{\rm LN} = 20$~kyr \citep{Halpern2014}. A sufficient number of particle injection events are considered, which are equally spaced along the source trajectory during this time, with $\Delta t$ shorter than the typical synchrotro and IC cooling timescales for particle Lorentz factors $\gamma_{\rm e}\lesssim \gamma_{\rm max}$. 
After injection, we consider electron diffusion in the surrounding ISM. A diffusion coefficient $D_{\rm diff}(E) = 3 \times 10^{27} \, \left(  \frac{E/\rm{GeV}} {B/\rm{ 3}\mu \rm{G}} \right)^{1/2}$ is assumed \citep{Gabici2007}. For simplicity, we assume that diffusion is isotropic from the injection point up to distances $l_{\rm diff} \sim (D_{\rm diff}\,t\,)^{1/2}$ (the medium is considered uniform, and thus adiabatic losses are not relevant). 

Freshly injected particles, with injection times $t_{\rm inj} \leq 1$~kyr, are presumably responsible for the X-ray emission observed from the jet-like features in the \lneb. For this injection period we consider distinct conditions of the ISM, which we label as the ``JET region'', characterized by a magnetic field $B_{\rm JET} = 5$--30~$\mu$G that can be significantly higher than $B_{\rm ISM} \sim 5 \mu$G under which electrons evolve after $t_{\rm inj} > 1$~kyr. The emission from particles injected until $t \sim 1$~kyr diffusing into this JET region, and the emission resulting from their later diffusion in the ISM region (for $t \gtrsim 1$~kyr), was computed separately. We emphasize however that the injected electron particle population is the same in the JET and ISM regions, only the value of the corresponding magnetic field for each region varies. The former are evolved until $t_{\rm inj} = 1$~kyr only, whereas the latter evolves from $t_{\rm inj} = 1$~kyr up to $t_{\rm inj} = 20$~kyr. 
The different conditions from the JET and ISM regions follow the assumption that the JET region corresponds to a temporary feature that relocates as the pulsar moves, and to which we assign a typical crossing timescale of $\sim 1$ kyr based on the JET apparent width and the pulsar velocity. The origin of the JET feature could be related to the pulsar motion pulling the threaded ISM magnetic field and, as the pulsar propagates, new ISM magnetic lines become entangled at the front of the PWN whereas the ``older'' ones relax and turn back into normal ISM magnetic field conditions.

Synchrotron emission is computed by employing the formulae in \cite{Pacholczyk1970} together with the analytical approximations in \citet{Aharonian2010}. The cumulative radio emission is evaluated throughout the area surrounding the source. Synthetic synchrotron emission maps are computed, which are then convolved with the beam of a radio telescope at GHz frequencies in order to obtain concrete observational predictions.

In this study we aim at predicting the radio emission produced in the surroundings of the sPWN, and therefore we do not attempt to model in detail the observed X-ray emission from the jet, including its morphological properties. In this regard, we only consider isotropic diffusion of particles injected in the JET and ISM regions. The effects that may result from anisotropies generated for example by the structured geometry of the underlying magnetic field, are therefore disregarded. Our estimates are based on assumed values for the injected kinetic energy, the energy distribution of the emitting particles, and on the ISM and JET regions' magnetic field values. We do not consider higher-energy radiation channels, although we note that gamma-ray halos around PWN have long been discussed (see e.g. \citealp{Aharonian1995}). In this regard, extended TeV emission has been recently reported from the PWN in Geminga (\citealp{Abeysekara2017}), which is consistent with free particle propagation in the ISM (see discussion in \citealp{Giacinti2020}). Finally, we focus in this study we focus on the outcomes that may be expected in the case of \IGR. A general model extendable to any other sPWN is not explored in this paper. We note also that the values adopted for the parameters describing the injected particle distribution can be seen as rather optimistic (e.g. $\Gamma $ could be harder than the adopted value of 2 and the low-energy cutoff $\gamma_{\rm min}^{ \rm cut}$ could be more restrictive than $10^3$). These, however, are relatively unconstrained parameters for the case of the source under study. At the same time, the adopted values are such that they can provide estimates of the expected radio emission that can then be contrasted with dedicated observations.

                                   \section{Results}
                                \label{section:results}
%           ************************************************************         

%======================================
%_____________________________________________________________________

%\begin{comment}
\begin{figure*}[t]
 \begin{subfigure}{.5\textwidth}
   \centering
   \includegraphics[width=\linewidth]{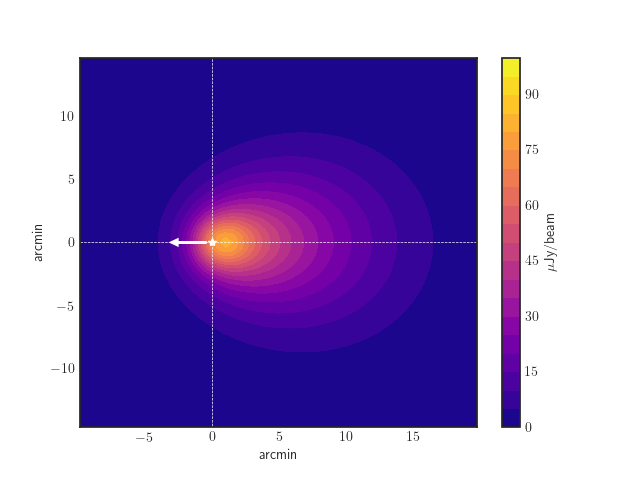}
   \end{subfigure}%
 \begin{subfigure}{.5\textwidth}
   \centering
   \includegraphics[width=\linewidth]{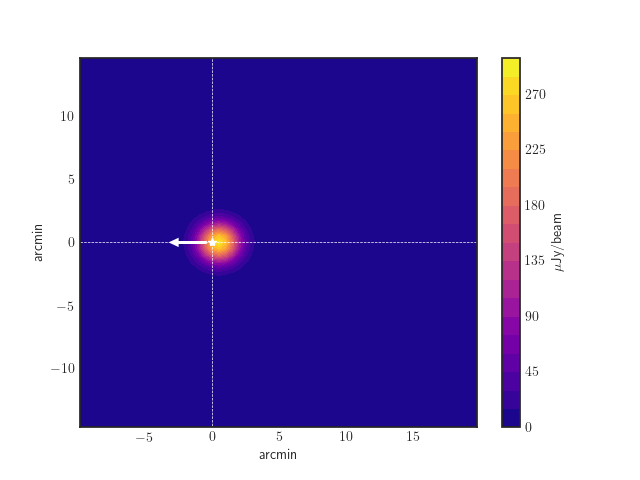}
   \end{subfigure}
  \caption
    {Synthetic synchrotron specific flux maps displaying the expected radio emission from \igr in the 0.9--1.7 GHz band in which MeerKAT operates, convolved with an instrument beam of $15\arcsec \times 15\arcsec$ (\citealp{Jonas2016}). The contribution from particles injected all along the \IGR's lifetime, which escaped the nebula and diffused into the ISM ($B_{\rm ISM} = 5\mu$G), is displayed in the left panel, and that from the JET region ($B_{\rm ISM} = 30\mu$G) is shown in the right panel. A white star indicates the position of the pulsar, which is assumed to move leftwards along the horizontal dashed-line. }
\label{figure:maps_gmin1e3}
\end{figure*}

%\end{comment}
%________________________________________________________________________
%======================================

\subsection{Injected particle distribution}

The energy distribution of particles injected along \IGR's lifetime is shown in Fig.~\ref{figure:particle_distribution}, which displays the two cases described in Sect.~\ref{section:model} for the value of $\gamma_{\rm min} = 10^{3}$ and $10^{5}$. For the high-energy cutoff, a value of $\gamma_{\rm max}^{ \rm cut} = 10^8$ is assumed in the two cases. Particles diffusing through the JET and the ISM regions are shown separately in Fig.~\ref{figure:particle_distribution}. For the ISM region, individual single-injection events every 1~kyr (dotted lines) are considered. The particles evolve under synchrotron and IC losses while accumulating, forming the final particle distribution (solid lines). The older the injections are, the lower their contribution is at the high-energy end of the final distribution. Freshly accelerated particles are considered only for the JET region (dashed lines in Fig.~\ref{figure:particle_distribution}).

\subsection{Estimated synchrotron fluxes}

The JET and the ISM regions differ in the assumed value for the magnetic field. We used $B_{\rm ISM} = 5 \mu$G to compute the synchrotron emission by each injection event and their total final contribution in the ISM. For the JET region, the value of $B_{\rm JET}$ ranges between that of the ISM and the value inferred to produce the observed X-ray emission in the Lighthouse Nebula (for electrons at a maximal energy $\gamma_{\rm max} \approx 10^8$) of $B_{\rm ISM} = 30 \mu$G \citep{Pavan2014}. The energy flux of the predicted synchrotron emission is displayed in Fig.~\ref{figure:SEDs} for the cases $\gamma_{\rm min}^{\rm cut} = 10^3$ and $10^5$. Individual contributions and total emission in the ISM are displayed in red with dashed and solid lines, respectively. The energy flux from the JET region is in blue, for five different values of $B_{\rm JET}$: 5, 10, 15, 20 and 30$\mu$G.

The radio emission for a low-energy cutoff $\gamma_{\rm min}^{ \rm cut} = 10^3$ reaches values of a few $\times 10^{29}$~erg~s$^{-1}$, which should be at levels detectable by existing facilities sensitive in this band. Radio emission is strongly suppressed for $\gamma_{\rm min}^{ \rm cut} = 10^5$, which instead provides a larger number of particles at higher energies. Current X-ray satellites can detect the JET region in X-rays even for modest magnetic fields $\sim 15 \mu$G. The total radio emission from the ISM region exceeds that of the JET region by a factor 2--10, depending on the value of $B_{\rm JET}$ and the adopted value of $\gamma_{\rm min}$. The JET region is however brighter than any single injection event in the ISM when taking a similar injection time lapse $\sim 1$~kyr. We include in Fig.~\ref{figure:SEDs} the observation bands of the radio telescope MeerKAT and the X-ray satellite \textit{Chandra}, noted by orange and green bands, respectively.

\subsection{Synthetic radio maps}

The cooled particle distribution injected in each event producing the synchrotron emission described above is distributed isotropically from its injection point. The synchrotron maps displayed in Fig.~\ref{figure:maps_gmin1e3} reflect this isotropy in the estimated spatial extension of the emission. The injection locations along the pulsar trajectory produce a ``trail'' of radio emission that is brighter along this direction, whereas the sum of the single events display an overall elliptical geometry encompassing the sPWN position. For the JET region, a single injection produces a radially symmetric extended source in our simplified model. In Fig.~\ref{figure:maps_gmin1e3} the maps display the angular distribution of the predicted synchrotron fluxes, which has been convolved with the beam of a radio telescope of $15\arcsec \times 15\arcsec$, such as that of the recently operative MeerKAT telescope (\citealp{Jonas2016}). We report only the case $\gamma_{\rm min}^{ \rm cut} = 10^3$ in Fig.~\ref{figure:maps_gmin1e3}, for which radio emission can reach detectable levels, above $\sim 50~ \mu$Jy~beam$^{-1}$. For $\gamma_{\rm min}^{ \rm cut} = 10^5$ radio emission would be undetectable. The spatial distribution in our model provides evidence that the JET region is the brightest spot around \IGR, but it is concentrated in a relatively small angular region of a few arcmins. We did not attempt to introduce further complexity with a nonisotropic particle distribution, which may produce extended structures in the radio maps, for example rectilinear jet-like features such as the ones reported in X-rays.

                                \section{Discussion}
                             \label{section:discussion}
%           ************************************************************         

The origin of the extended jet-like features observed in the Lighthouse Nebula and other sPWNs is unclear. 
The high velocity of these systems and the apparent rectilinear geometry of their associated outflows, with no appreciable signature of bending as expected due to the ambient medium ram pressure, suggest alternative mechanisms responsible for such jet-like features. \cite{Bandiera2008} proposed a scenario in which the jets could be produced by high-energy electrons escaping the parent PWN and diffusing into the ambient magnetic field. Recent 3D magnetohydrodynamic (MHD) numerical simulations seem to support such a possibility \citep{Barkov2019a, Bucciantini2020}. Here we provide further observational predictions to probe this hypothesis in the context of the \LNEB, modeling the radio emission from the immediate surroundings of the source.

The results reported in Sect.~\ref{section:results} indicate that, for suitable parameters describing the injected particle energy distributions escaping the sPWN, substantial radio emission around \igr could be produced by these particles when diffusing into the ISM. We note that the value assumed for these parameters (e.g. $\Gamma = 2$ and $\gamma_{\rm min}^{ \rm cut} = 10^{3}$) were chosen so as to account for cases in which detectable radio fluxes may be expected, whereas in others (e.g. $\gamma_{\rm min}^{ \rm cut} = 10^{5}$) such emission would be suppressed. For the former case, the JET region provides the larger contribution, since $B_{\rm JET}$ is larger than $B_{\rm ISM}$. The fact that $B_{\rm JET}>B_{\rm ISM}$ could follow the presence of streaming instabilities generated in the flow of relativistic electrons, generating a turbulent component of the magnetic field responsible for the field amplification, and for the confinement of the radiating electrons themselves \citep{Bykov2017}. The total cumulative radio emission from the more diffuse ISM region, on the other hand, can be a factor of a few to $\sim$20 times higher than that in the JET region. According to our model, observations should therefore reveal a bright concentration toward the jet region, and a broader, dimmer diffuse radio nebular emission surrounding the source. Compared to the radio emission from the PWN tail reported in \citep{Pavan2014}, the emission from the ISM region in our model displays somewhat softer fluxes. The nebular emission in the ISM region reported here might be intrinsically softer than that within the PWN in \IGR. Otherwise, the presence of several populations of accelerated electrons in the PWN particle injection spectrum (see e.g. \citealp{Atoyan1996}), could also make the radio fluxes harder. For the JET region, although we do not aim to describe the morphological and spectral properties of the extended jet-like features observed in the Lighthouse Nebula in X-rays, our model predicts fluxes compatible with the ones observed at this band, at a level $L_{\rm 2-10keV}\gtrsim 10^{33}$~erg~s$^{-1}$, when assuming previously derived values of $B_{\rm JET}$ of a few $\times$ 10~$\mu$G \citep{Pavan2014}. At higher energies, in the gamma-ray band, it is worth noting that, given the value of $U_{\rm IR}$ the produced IC gamma-ray luminosity by the electrons responsible for the JET X-rays should be a factor $\sim U_{\rm IR}/(B_{\rm JET}^{2}/8\pi) \sim 0.3$ times that observed in X-rays, $L_{\rm 2-10keV}$ (a detailed treatment is left for future work).

As stated above the radio emission predicted in our model strongly depends on the assumed limits in the energy distribution of the emitting particles. As displayed in Fig.~\ref{figure:SEDs}, right panel, a strong low-energy cutof of $\gamma_{\rm min}^{\rm cut} \gtrsim 10^{5}$ would render these structures too dim to be detected. On the contrary, the synchrotron radio flux maps displayed in Fig.~\ref{figure:maps_gmin1e3} for $\gamma_{\rm min}^{\rm cut} \gtrsim 10^{3}$ indicate that extended structures should be resolvable by current radio telescopes capable of reaching sensitivities at the level of $\gtrsim 50 \mu$Jy/beam. The value of $\gamma_{\rm min}^{\rm cut}$ depends on the mechanisms through which accelerated particles are able to escape the PWN region. For this to occur, the diffusion distance within the bow-shock region should be larger than the shocked wind layer thickness. Given that this is a  several times smaller than the bow-shock size, the escape energy is just a factor order of 10 smaller than that given by the Hillas limit, which for the \lneb has an associated Lorentz factor of $\gtrsim 10^8$.
Nevertheless, particles with lower energies may still be able to escape the PWN along the expanding PWN tail where the magnetic field density decreases and/or the geometry of the shock interface with the ambient ISM becomes more favorable for electrons to diffuse away. In this regard, it is worth noting that for the case of the \LNEB, the base of the main jet-like feature seems to be shifted to a position behind the pulsar with respect to the incoming medium (see e.g. \citealp{Pavan2016}; see also the inset in Fig.~9 in \citealp{Kargaltsev2017}), where the PWN conditions could imply lower values for the particles' Lorentz factor to escape the nebula. Alternatively, particle escape can follow magnetic reconnection between the sPWN and the ISM magnetic fields (see e.g. \citealp{Barkov2019, Bucciantini2020}), as suggested in the case of the rectilinear structures detected in J1809-1917, for which the gyroradius of the X-ray emitting electrons is by far shorter than the typical size of the nebula (\citealp{Klingler2020}).

Finally, although in this report we have focused on the \lneb, we note that other sPWNe may also display extended radio emission from electrons escaping their nebulae. If confirmed, the detection of these radio structures would represent a footprint for an efficient particle escape mechanism taking place in these systems, allowing in turn the formation of extended, rectilinear, and misaligned jet-like structures as observed in some of the most extreme sPWNe.

\begin{acknowledgements}
PB, VBR, JMP and XZ acknowledge the financial support by the Spanish Ministerio de Econom\'{i}a, Industria y Competitividad (MINEICO/FEDER, UE) under grant AYA2016-76012-C3-1-P, from the State Agency for Research of the Spanish Ministry of Science and Innovation under grant PID2019-105510GB-C31 and through the “Unit of Excellence Mar\'{i}a de Maeztu 2020-2023” award to the Institute of Cosmos Sciences (CEX2019-000918-M), and by the Catalan DEC grant 2017 SGR 643.  
\end{acknowledgements}

%\newpage
\bibliographystyle{aa} 
\bibliography{bibliography} 
\end{document}